\documentclass[useAMS,usegraphicx]{mn2e}
\usepackage{rotate}
\usepackage{times}
\newif\ifAMStwofonts
\AMStwofontstrue

\title[X-ray Spectroscopy of Bright ULXs]
      {
On The Robustness of Cool Disc Components in Bright ULXs
      }

\author[J. M. Miller et al.]
       {J. M. Miller,$^1$ 
	A. C. Fabian,$^2$ 
	and M. C. Miller$^3$\\
$^{1}$ Department of Astronomy, University of Michigan, 500 Church
Street, Ann Arbor, MI 48109, USA, jonmm@umich.edu \\
$^{2}$ Institute of Astronomy, University of Cambridge, Madingley
Road, Cambridge CB3 0HA\\
$^{3}$ Department of Astronomy, University of Maryland, College Park,
MD, 20742, USA\\
}
\date{Accepted. Received. }
\pagerange{\pageref{firstpage}--\pageref{lastpage}}
\pubyear{2005}
\begin{document}
\maketitle
\label{firstpage}

\begin{abstract}
In this letter, we comment on the robustness of putative cool ($kT
\simeq 0.2$~keV) accretion disc components in the X-ray spectra of the
most luminous ($L_{X} \simeq 10^{40}$~erg/s) ultra-luminous X-ray
sources (ULXs) in nearby normal galaxies.  When compared to
stellar-mass black holes, the low disc temperatures observed in some
ULXs may imply intermediate-mass black hole primaries.  Recent work
has claimed that such soft excesses are unlikely to be actual disc
components, based on the lack of variability in these components, and
in the overall source flux.  Other work has proposed that alternative
phenomenological models, and complex Comptonisation models, rule-out
cool disc components in ULX spectra.  An inspection of the literature
on Galactic stellar-mass black holes and black hole candidates
demonstrates that the flux behaviours seen in specific ULXs are
consistent with phenomena observed in well-known Galactic X-ray
binaries.  Applying Comptonisation models to simulated disc blackbody
plus power-law spectra shows that at the sensitivity achieved in even
the best ULX spectra, Comptonisation fits are highly model-dependent,
and do not yield meaningful constraints on the accretion flow.  In
contrast, the need for a soft, thermal component does not appear to be
model-dependent.  As we have previously noted, soft thermal components
in ULX spectra may not represent accretion discs, but present
alternatives to this interpretation are not robust.
\end{abstract}

\begin{keywords}
galaxies: normal -- 
X-ray: galaxies 
\end{keywords}


\section{Introduction}

The nature of ultra-luminous X-ray sources (ULXs, $L_{X} \geq 2\times
10^{39}$~erg/s) in nearby normal galaxies has been a topic of great
interest in the {\it Chandra} and {\it XMM-Newton} era.  The apparent
luminosity of these sources can exceed the isotropic Eddington limit
for a $10~M_{\odot}$ black hole, leading to the possibility that some
ULXs may harbour intermediate-mass black holes ($10^{2-5}~M_{\odot}$).
Initially, this debate was framed in absolute terms, e.g., What is the
nature of the ULX phenomenon?  Such questions were likely posed in an
erroneously simple manner.  Every new class of sources is soon divided
into subclasses with further study, and is found to be a broad
designation covering a heterogeneous group of sources.  The fact that
Galactic stellar-mass black holes can appear to be mildly
super-Eddington (see, e.g., McClintock \& Remillard 2005) may argue
that many ULXs are stellar-mass black holes.  However, a growing
subset (fewer than 10 presently) is emerging that may represent a
class of intermediate mass black holes.

The lessons learnt from decades of X-ray spectral and timing studies
of Galactic stellar-mass black hole binaries and black hole candidates
would seem to indicate that a small number of the most luminous ULXs
($L_{X} \simeq 10^{40}$~erg/s, and above) may harbour black holes of a
few hundred or few thousand solar masses (see, e.g., Colbert \&
Mushotzky 1999; Strohmayer \& Mushotzky 2003; Miller et al.\ 2003;
Cropper et al.\ 2004; Miller, Fabian, \& Miller 2004a; Kaaret, Ward,
\& Zezas 2004; Miller \& Colbert 2004).  Intermediate mass black holes
are implied in these sources via Eddington luminosity scaling, and/or
multi-wavelength properties that argue against beaming, and/or scaling
characteristic frequencies found in the X-ray flux, and/or scaling
apparent inner disc temperatures in the X-ray spectra.

In the case of inner disc temperatures, intermediate mass black holes
may be implied because the temperatures implied ($kT \simeq 0.2$~keV)
are well below the temperatures typically measured in stellar-mass
black holes accreting near their Eddington limits ($kT \simeq
1-2$~keV), and $T \propto M^{-1/4}$ for standard black hole accretion
discs (see, e.g., Miller, Fabian, \& Miller 2004a).  However, the
robustness of these soft components has recently been
questioned.  It has been suggested that the variability properties of
ULXs and the soft components in particular, argue against associating
them with discs (e.g. Dewangan, Griffiths, \& Rao 2005; Goad et al.\
2005).  Other work has suggested that alternative spectral models may
be more appropriate, and may rule-out the possibility of cool discs
and intermediate mass black holes in the most luminous ULXs (e.g. Goad
et al.\ 2005, Roberts et al.\ 2005).  In this letter, we examine these
arguments in detail, in the context of better-understood
Galactic stellar-mass black hole binaries and AGN.

\section{On Disc Temperature and Flux Variations}

While some broad expectations can be derived from robust theoretical
considerations, the detailed phenomenology of accretion discs and
their inner workings are poorly understood at present.  One such simple
expectation is that disc temperature should directly correlate with
disc flux, assuming apparent disc flux is a good proxy of the mass
accretion rate through the disc.  Dewangan, Griffiths, \& Rao (2005)
have recently claimed that the cool, soft component in the X-ray
spectrum of NGC 1313 X-1 may not actually be an accretion disc because
``its blackbody temperature is similar in three XMM-Newton
observations, despite a change in the observed flux by a factor of
about two.''  In the absence of a detailed theoretical understanding
of accretion discs, insight may be gained by examining the observed
behaviour of stellar-mass black holes and black hole candidates.  In
these systems, both the presence of accretion discs and their
manifestation in X-ray spectra are beyond doubt at high mass accretion
rates.

The temperatures measured in NGC 1313 X-1 by Dewangan, Griffiths, \&
Rao (2005) range between kT$=0.14-0.21$~keV, depending on the model
and spectrum.  In fits which apply the same soft component model to
different source flux levels, the measured temperature variations
range between 5--20\%.  The rich literature on the behavior of
Galactic black holes clearly shows that disc components in these
better-understood systems can manifest exactly such behaviour.  In the
Galactic black hole 4U 1543$-$47, Park et al. (2004) report
temperature variations as small as 6\% over a factor of 2.1 in disc
flux (obs.\ 18 versus obs.\ 24).  In XTE~J1550$-$564, Sobczak et
al. (2000) report disc temperature variations as small as 4\% over a
factor of 2.8 in disc flux (obs.\ 152 versus obs.\ 156).  In
GRO~J1655$-$40, Sobczak et al.\ (1999) show that the disc temperature
can differ by as little as 5\% over a factor of more than 4.0 in disc
flux (1996 Oct.\ 27 versus Nov.\ 02).  In 4U~1630$-$472, Trudolyubov,
Borozdin, \& Priedhorsky (2001) show that the disc temperature can
vary by as little as 1\% across a factor of 2.9 in disc flux (obs.\ 3
versus obs.\ 31).  Finally, in XTE~J1748$-$288, Miller et al. (2001)
report a temperature variation of only 2\% across a factor of 5.3 in
disc flux (obs.\ 8 versus obs.\ 9).

Clearly, based on a comparison with Galactic black holes, the fact
that a temperature--flux relation does not hold at all times does not
preclude associating a soft thermal component with an accretion disc.

\section{On Fast X-ray Variability}

Goad et al.\ (2005) claim that the absence of fast variability in
the broad-band X-ray flux of Ho II X-1 argues against the presence of
an intermediate mass black hole, and against the possibility that the
soft excess in this ULX arises from a cool accretion disc.  This
argument is based on the premise that low fractional variability is
not observed at low fractional Eddington accretion rates in
stellar-mass black holes and black hole candidates.  However, existing
literature shows that this is incorrect.  Smith et al. (2001) report
that the fractional variability in GRS~1758$-$258 dropped to values
consistent with zero, in a low-flux state (approximately
0.04~$L_{Edd.}$ for a $10~M_{\odot}$ black hole at $d=8.5$~kpc)
dominated by cool disc emission.  In an observation of 4U 1957$+$11 at
0.05~$L_{Edd.}$ (for a $10~M_{\odot}$ black hole at $d=7$~kpc),
Wijnands, Miller, \& van der Klis (2002) report a total fractional
variability of only 1.4\% where the hard component is still 88\% of
the total flux (see obs.\ 9).  

Clearly, low fractional variability is possible at low flux levels in
stellar-mass black holes and black hole candidates.  We note that
episodes of low fractional variability are not frequent in
stellar-mass black holes and black hole candidates, but they are
observed, and less extreme examples are more common.  Thus, low
fractional variability cannot be used to argue against applying the
same spectral interpretations to ULX spectra that are used in
stellar-mass black holes and black hole candidates.

\section{On Phenomenological Spectral Models}

It is common to model the spectra of accretion-powered sources with
simple, phenomenological models, usually consisting of a thermal disc
component at low energy and a hard (sometimes non-thermal)
power-law-like component at high energy.  Models of this sort reflect
the basic expectation that a thermal accretion disc, and hard emission
from the inverse Compton scattering of disc photons, should be the
dominant sources of emission in accreting black holes.  It has been
claimed that the roles of these components can be flipped in the
spectra of some ULXs; that is, the low energy emission might be
dominated by a steep power-law component, and the hard emission might
be dominated by a very hot blackbody component (e.g. Roberts et al.\
2005).  While such fits may be statistically permissible in spectra of
poor sensitivity, and/or in dipping phases, it is worth examining
whether or not there is any physical justification or any precedent
for such a model in ordinary phases.

A component arising from the inverse-Compton scattering of disc
photons cannot dominate the disc flux below the peak of the blackbody
distribution, so an alternative explanation for the hard component is
required.  It is possible that the hard component is partly or even
mostly due to synchrotron emission from a jet (e.g. Markoff, Fender,
\& Falcke 2001).  In this case, (depending partially on whether or not
the jet power-law has a break) the power-law distribution might extend
to low energy and dominate both below and above the disc blackbody
distribution.  This scenario might broadly correspond to one in which
cool thermal and hard non-thermal components could be flipped.
However, jets are only predicted to contribute significantly to the
X-ray flux in the ``low/hard'' state; if ULXs with implied
luminosities near $10^{40}$~erg/s are in a low/hard state, then they
are accreting at a small fraction of their Eddington limits, and
intermediate-mass black holes would still be implied by their
luminosity.  Moreover, in such states, hard power-law components are
typically observed; $\Gamma = 1.5-1.7$ is common.  The power-laws
required in flipped spectral models of ULXs are extremely soft, with
$\Gamma = 3-4$.  There is no evidence for steady jet production in
stellar-mass black holes and black hole candidates, in states with
such steep power-law indices.  Given that such models are not
dramatically better fits than standard models and that
better-understood black holes do not manifest the behaviour implied, it
seems likely these models only reflect statistical (not physical)
possibilities.

\section{Complex Comptonisation Models: the Echoes of Intention}

The power of simple phenomenological spectral models is that they can
be tightly constrained, even when fitting spectra of limited
sensitivity.  Compared to the best AGN and stellar-mass black hole
spectra, present ULX spectra are certainly of that type.  More
physically-motivated models, many of which attempt to e.g. properly
calculate the up-scattering of disc X-rays in a corona, predict subtle
curvature which is difficult to detect in low signal-to-noise spectra.
Moreover, such models come with a price: additional parameters.  A
model consisting of disc blackbody and power-law components, modified
by line of sight absorption, has 5 parameters.  Replacing the
power-law component with a simple Comptonisation model -- ``compTT''
-- drives the number of model parameters up to 9.  A model which
employs a full code such as ``eqpair'' (Coppi 2002), which includes a
disc distribution, has 21 model parameters after adding one parameter
to account for line-of-sight absorption.  At the signal-to-noise
typical of present ULX spectra, the parameters of complex models
cannot be regarded as reliable.  The resultant fits will (by force)
reflect the assumptions of the model and biases of the modeler in such
circumstances.  Moreover, resultant fits are likely to be
statistically degenerate with fits obtained using the same model with
different bounds or choices for given parameters.

In order to illustrate that even simple Comptonisation models can not
only yield degenerate results based only on assumptions, but can
actually yield incorrect ``constraints'' on the accretion flow, we
have simulated a disc blackbody plus power-law spectrum and fit
it with a basic Comptonisation model.  The power of this approach is
that the nature of the spectrum is known a priori, and the ability of
complex models to distort an intrinsically simple spectrum or provide
false information is effectively tested.  

The simulated spectral parameters were taken from disc blackbody plus
power-law fits made to NGC 1313 X-1 and reported in Miller, Fabian, \&
Miller (2004b).  The parameters for the simulated spectrum are as
follows: $N_{H} = 3.1\times 10^{21}~{\rm cm}^{-2}$, $kT_{dbb} =
0.23$~keV, $N_{dbb} = 28$, $\Gamma = 1.76$, $N_{pl} = 4.9\times
10^{-4}~{\rm ph}~{\rm cm}^{-2}~{\rm s}^{-1}$.  The same {\it
XMM-Newton}/EPIC-pn ``primefullwindow'' mode (plus medium optical
blocking filter) response files were used to generate a simulated
100~ksec EPIC-pn spectrum, using the XSPEC task ``fakeit''.  Note that
this simulated observation is as long as the longest single
observation of any ULX obtained with {\it XMM-Newton}.  Only Poisson
noise is included in simulating a spectrum in this way; in generating
and fitting the simulated spectrum, we considered no additional noise
or background.  In effect, then, the simulated spectrum is of a higher
sensitivity than any ULX spectrum yet obtained.

Prior to fitting, the simulated spectrum was grouped to require 10
counts/bin using the FTOOL ``grppha'' to ensure the accuracy of
$\chi^{2}$ fitting results.  Spectral fits were made in the
0.3--10.0~keV band using XSPEC 11.3.2.  Reported errors on fit
parameters are 90\% confidence errors for one parameter of interest.
The simple Comptonisation model used consisted of a disc blackbody
component, and the ``compTT'' thermal Comptonisation model, modified
by line-of-sight absorption.

A statistically acceptable fit to the simulated spectrum can be
obtained with a low-temperature, optically-thick corona similar to the
fits spectra of Ho II X-1 reported by Goad et al.\ (2005).  Of course,
in the simulated spectrum, there is no component which directly
corresponds to such a low-temperature, optically-thick corona.  With
this model, we obtain the following parameter values: $N_{H} = 2.8(3)
\times 10^{21}~{\rm cm}^{-2}$, $kT_{dbb} = 0.23(3)$~keV, $N_{dbb} =
20(4)$, $T_{seed} = 0.41(3)$~keV, $kT_{e} = 6.8(3)$~keV, $\tau =
4.1(2)$, and $N_{compTT} = 9.3(2)\times 10^{-5}$, and a formally
acceptable fit statistic of $\chi^{2}/dof = 1226/1245$.  Let us refer
to this model as ``CM1''.  In this fit, we fixed the redshift
parameter in ``compTT'' to be equal to zero, and the approximation
flag equal to unity (corresponding to a disc plus corona geometry).

It is also possible to obtain a statistically acceptable fit to the
simulated spectrum, with a hot, optically-thin corona.  Comptonisation
in a hot corona should produce less spectral curvature, and should
better correspond to a simple power-law, though we again note that
there is of course no Comptonisation component in the simulated
spectrum.  With this model, we obtain the following parameter values:
$N_{H} = 2.7(3) \times 10^{21}~{\rm cm}^{-2}$, $kT_{dbb} =
0.22(2)$~keV, $N_{dbb} = 33(4)$, $T_{seed} = 0.22(3)$~keV, $kT_{e} =
49(3)$~keV, $\tau = 0.83(3)$, and $N_{compTT} = 2.1(1)\times 10^{-5}$,
and a formally acceptable fit statistic of $\chi^{2}/dof = 1228/1245$.
Let us refer to this model as ``CM2''.  The incredible similarity of
this spectral model to that detailed above, despite very different
model parameters and a very different physical picture, is depicted in
Figure 1.

Statistically acceptable fits can also be achieved using  broken
power-law or exponentially cut-off power-law models, though the
simulated power-law contains neither a break nor a cut-off.  Using a
broken power-law, the following parameters were obtained: $N_{H} =
2.7(2) \times 10^{21}~{\rm cm}^{-2}$, $kT_{dbb} = 0.29(2)$, $N_{dbb} =
7.3(9)$, $\Gamma_{1} = 1.71(2)$, $E_{br} = 5^{+2}_{-1}$~keV,
$\Gamma_{2} = 1.82(7)$, $N_{pl} = 4.3(1) \times 10^{-4}~{\rm ph}~{\rm
cm}^{-2}~{\rm s}^{-1}$, and $\chi^{2}/\nu = 1220/1258$.  We note that
a break with $\delta(\Gamma) = 0.3$ can also be accommodated,
statistically.  Using an exponentially cut-off power-law, the
following parameters are obtained: $N_{H} = 2.3(2) \times 10^{21}~{\rm
cm}^{-2}$, $kT_{dbb} = 0.34(3)$, $N_{dbb} = 4.3(8)$, $\Gamma_{1} =
1.1(1)$, $E_{cut} = 7.7(3)$~keV, $N_{pl} = 3.1(1) \times 10^{-4}~{\rm
ph}~{\rm cm}^{-2}~{\rm s}^{-1}$, and $\chi^{2}/\nu = 1228/1259$.
Breaks and cut-offs are merely possible, not required, in present ULX spectra.

We also simulated 100~ksec spectra, based on the parameters
obtained in CM1 and CM2.  The simulated spectrum based on CM1 can be
fitted with a model that has the following parameters: $N_{H} = 2.6(2)
\times 10^{21}~{\rm cm}^{-2}$, $kT_{dbb} = 0.12(2)$~keV, $N_{dbb} =
110(60)$, $T_{seed} = 0.14(2)$~keV, $kT_{e} = 53(3)$~keV, $\tau =
0.89(4)$, and $N_{compTT} = 2.3(2)\times 10^{-5}$, yielding an
acceptable fit statistic of $\chi^{2}/dof = 1307/1245$.  The simulated
spectrum based on CM2 can be fitted with the following model
parameters: $N_{H} = 2.2(2) \times 10^{21}~{\rm cm}^{-2}$, $kT_{dbb} =
0.28(3)$~keV, $N_{dbb} = 5.2(9)$, $T_{seed} = 0.17(2)$~keV, $kT_{e} =
6.0(2)$~keV, $\tau = 4.2(2)$, and $N_{compTT} = 1.8(2)\times 10^{-4}$,
yielding an acceptable fit statistic of $\chi^{2}/dof = 1094/1245$.
That is, at the sensitivity of the best present ULX spectra, a cool
optically-thick coronal spectrum can be fitted with a hot,
optically-thin model, and vice versa. 

Clearly, even simple Comptonisation models cannot be used to infer the
nature of the corona in present ULX spectra.  Although the nature of
the corona cannot be inferred reliably, the results detailed above
show that the assumed nature of a hard spectral component does not
strongly affect the need for a soft component, or the temperatures
derived when this component is fit with a disc model.  This is true
when working with real data as well: Goad et al.\ (2005) require a
cool disc to obtain a good fit to the spectrum of Ho~II X-1, though
its significance is dismissed based on the Comptonisation component
fit to that spectrum.

Whereas one might expect that the parameters of Comptonizaton models
merely cannot be constrained reliably at the signal-to-noise achieved
in ULX spectra, the true situation is in fact much worse: even simple
Comptonisation models can easily be manipulated to give the answer the
observer wants, and that answer can in fact be entirely incorrect
though statistically acceptable.  This finding calls into
question the validity of the models considered and the utility of
these and yet more sophisticated models (e.g. eqpair, see Roberts et
al.\ 2005) in the ULX regime.

Phenomenological spectral models, while perhaps physically naive, are
robust and useful in that their parameters can be constrained
reliably.  Fits with such models are reproducible, falsifiable, and
therefore meaningful.  They provide a common currency through which
different sources can be compared consistently.  Complex models which
are not statistically required, and which are highly malleable, cannot
be falsified, and therefore necessarily fail to provide scientific
insights.

\section{Summary and Conclusions}

We have shown that a variety of arguments against interpreting
apparent cool thermal components in the spectra of very luminous ULXs
are contradicted by published results from stellar mass black holes
and black hole candidates, by simple physical considerations, and by
critically examining the ability of complex models to yield meaningful
results in low signal-to-noise regimes.  The absence of direct
correlations between apparent disc temperatures and fluxes observed in
some ULXs is a phenomenon commonly observed in stellar-mass black
holes and black hole candidates.  Moreover, limited fast X-ray
variability has been observed in stellar-mass black holes and black
hole candidates, even in hard phases at low fractional Eddington
luminosities.  Spectroscopy of stellar-mass black holes and physical
considerations strongly argue for associating thermal distributions
with the low energy portion of a spectrum, and scattered or
non-thermal distributions with the high energy portion.  Finally, we
have shown that even at the sensitivity achieved in the best present
ULX spectra, complex Comptonisation models can yield false inferences
and give an incorrect picture of the accretion flow geometry.

In the case of NGC 1313 X-1 and M81 X-9, present spectra have been
sufficient to show that cool disc plus power-law models are at least
5$\sigma$ better than single-component models, broken power-law
models, and models with significantly sub-solar abundances in gas
along the line of sight (Miller et al.\ 2003; Miller, Fabian, \&
Miller 2004b).  The same work has shown that cool discs are robust
against specific choices of disc models and specific choices of hard
component models.  Even in cases where Comptonisation models are
invoked to argue against interpreting soft excesses as discs, cool
disc components are required to achieve acceptable fits (e.g. Goad et
al.\ 2005).  This further highlights the robustness of cool thermal
components, and the perils of Comptonisation models in low
signal-to-noise spectra.

The cool disc interpretation is merely the most plausible one, based
on a comparison to other black holes, and principally stellar-mass
black holes.  As we have previously noted, however, cool thermal
components in the spectra of very luminous ULXs are not necessarily
disc components, and not all soft components in the spectra of
accreting black holes are well-understood.  The soft excess observed
in some Seyfert-1 galaxies is too hot to easily be attributed to a
disc, and its nature is uncertain.  Miller, Fabian, \& Miller (2004a)
noted this aspect of Seyfert-1 spectra, in anticipation of the
possibility that some of the ULXs which have cool thermal components
may actually be background AGN.  Indeed, this proved to be the case
for the ULX Antennae X-37 (Clark et al.\ 2005).  It is possible that
the soft excess in genuine ULXs is similar to that in some Seyfert-1
spectra.  However, if both phenomena are due to
relativistically-blurred disc emission lines (Crummy et al.\ 2005),
then the soft component in ULXs indicates that the emission is
isotropic and intermediate-mass black holes may still be required via
simple Eddington scaling arguments.  We note that a preliminary
investigation we have undertaken reveals that {\it XMM-Newton} spectra
of ULXs with apparent cool thermal components can also be fit
acceptably with blurred disc reflection models.

While this letter demonstrates that cool thermal components are likely
more robust than than inferences derived from models for the hard
component in ULX spectra, it also serves to demonstrate that present
ULX spectra are intrinsically a low signal-to-noise regime.  All
spectral fits, whether simple or complex, must be regarded cautiously
in such circumstances, but the most robust conclusions probably derive
from simple models that can be constrained by the data.  Very deep
observations of ULXs with {\it XMM-Newton} (350~ksec or longer) may be
the only means of parsing the nature of soft and hard components in
very luminous ULXs in more detail.


\section*{Acknowledgements}

We thank Jeroen Homan, Joel Bregman, and Jimmy Irwin for useful
discussions.  MCM was supported in part by NASA ATP grant 5-13229, and
by a senior NRC fellowship during a sabbatical at the Goddard Space
Flight Center.


\begin{figure*}
\scalebox{1.0}{\includegraphics[angle=0]{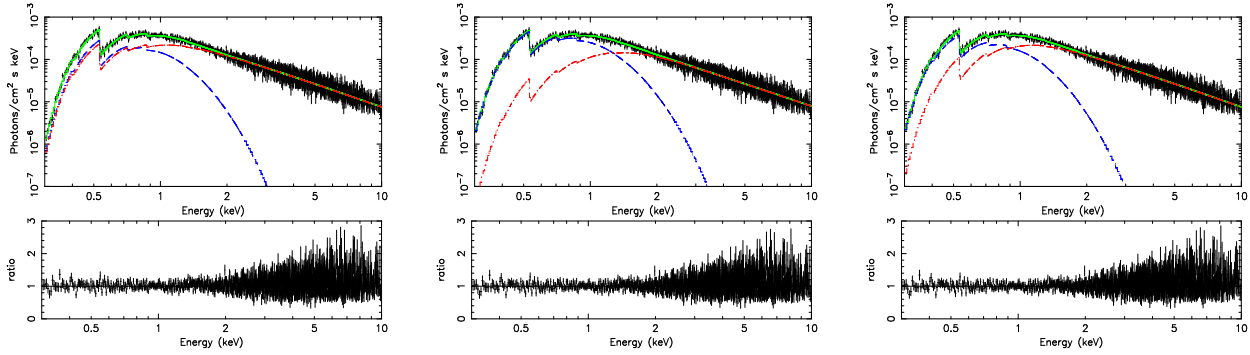}}
\caption
{The figure above shows the results of model fits to a simulated
spectrum consisting of disc blackbody and power-law components.  The
fit shown in the left panel was obtained using disc blackbody and
power-law components.  The fit shown in the middle panel was obtained
using a model consisting of a disc blackbody and a cool ($kT =
6.8$~keV), optically-thick ($\tau = 4.1$) Comptonising coronal
component. The fit shown in the right panel was obtained using a model
consisting of a disc blackbody and a hot ($kT = 48$~keV),
optically-thin ($\tau = 0.8$) Comptonising coronal component.
Although the fits are statistically equivalent, it is clear that very
different Comptonisation models produce only minor changes in the
0.5--10.0~keV band, and both Comptonisation models yield false
implications about the simulated spectrum.}
\end{figure*}

\end{document}